\documentclass{aa}  
\usepackage{graphicx}
\usepackage{txfonts}
\usepackage{longtable}
\usepackage{natbib}
\bibpunct{(}{)}{;}{a}{}{,}
\def\sOri{{$\sigma$Orionis}}
\begin{document}
   \title{The mass function of IC\,4665 revisited by the UKIDSS Galactic Clusters Survey \thanks{This work is based in part on data obtained as part of the UKIRT Infrared Deep Sky Survey. The United Kingdom Infrared Telescope is operated by the Joint Astronomy Centre on behalf of the Science and Technology Facilities Council of the U.K.} \thanks{This work is partly based on observations obtained at the Canada-France-Hawaii Telescope (CFHT), which is operated by the National Research Council of Canada, the Institut National des Sciences de l'Univers of the Centre National de la Recherche Scientifique of France,  and the University of Hawaii.}\thanks{Tables A.1, B.1, and C.1 are only available in 
electronic form at the CDS via anonymous ftp to cdsarc.u-strasbg.fr 
(130.79.128.5) or via http://cdsweb.u-strasbg.fr/cgi-bin/qcat?J/A$+$A/
}}

   \subtitle{}

   \author{N. Lodieu \inst{1,2}
          \and
          W.-J. de Wit \inst{3}
          \and
          G. Carraro \inst{3}
          \and
          E. Moraux \inst{4}
          \and
          J. Bouvier \inst{4}
          \and
          N. C. Hambly \inst{5}
          }

   \offprints{N. Lodieu}

   \institute{Instituto de Astrof\'isica de Canarias, C/ V\'ia L\'actea s/n,
              E-38200 La Laguna, Tenerife, Spain\\
             \email{nlodieu@iac.es}
             \and
             Departamento de Astrof\'isica, Universidad de La Laguna (ULL),
             E-38205 La Laguna, Tenerife, Spain
             \and
             European Southern Observatory, Alonso de Cordoba 3107, Casilla
             19001 Santiago 19, Chile \\
             \email{gcarraro@eso.org, wdewit@eso.org}
             \and
             UJF-Grenoble 1/CNRS-INSU, Institut de Plan\'etologie et 
d'Astrophysique de Grenoble (IPAG) UMR 5274, Grenoble, F-38041, France \\
             \email{Estelle.Moraux@obs.ujf-grenoble.fr, Jerome.Bouvier@obs.ujf-grenoble.fr}
             \and
             Scottish Universities' Physics Alliance (SUPA),
             Institute for Astronomy, School of Physics \& Astronomy, University of Edinburgh,
             Royal Observatory, Blackford Hill, Edinburgh EH9 3HJ, UK \\
             \email{nch@roe.ac.uk}
             }

   \date{\today{}; \today{}}

 
  \abstract
   {Knowledge of the mass function in open clusters constitutes one 
way to constrain the formation of low-mass stars and brown dwarfs as
does the knowledge of the frequency of multiple systems and the properties 
of disks.}
   {The aim of the project is to determine the shape of the mass 
function in the low-mass and substellar regimes in the pre-main sequence 
(27 Myr) cluster IC\,4665, which is located at 350 pc from the Sun.}
   {We have cross-matched the near-infrared photometric data from the 
Eighth Data Release (DR8) of the UKIRT Infrared Deep Sky Survey (UKIDSS) 
Galactic Clusters Survey (GCS) with previous optical data obtained with
the Canada-France-Hawaii (CFH) wide-field camera to improve the 
determination of the luminosity and mass functions in the low-mass and 
substellar regimes.}
   {The availability of $i$ and $z$ photometry taken with the CFH12K camera
on the Canada France Hawaii Telescope added strong constraints to the UKIDSS
photometric selection in this cluster, which is located in a dense region of 
our Galaxy. We have derived the luminosity and mass functions of the cluster
down to $J$ = 18.5 mag, corresponding to masses of $\sim$0.025 M$_{\odot}$
at the distance and age of IC\,4665 according to theoretical models.
In addition, we have extracted new candidate members down to 
$\sim$20 Jupiter masses in a previously unstudied region of the cluster.}
   {We have derived the mass function over the 0.6--0.04 M$_{\odot}$
mass range and found that it is best represented by a log-normal
function with a peak at 0.25--0.16 M$_{\odot}$, consistent with the
determination in the Pleiades.}


   \titlerunning{The mass function of IC\,4665 revisited by the UKIDSS GCS}
   \maketitle
%

%
%
\section{Introduction}

The initial mass function (IMF), the number of stars formed in the
galactic disk and in clusters, was first defined by \citet{salpeter55}.
Later, \citet{miller79} and \citet{scalo86} conducted a more accurate 
study of the IMF. It represents
an important link between stellar and galactic evolution and is one of the
key tools for our understanding of the theories of star formation in
the high-mass, low-mass, and substellar regimes. The current knowledge
drawn from the latest sets of data available in the solar neighbourhood
and in clusters points towards a power-law or log-normal
representation of the IMF \citep{kroupa02,chabrier03,larson05}.

Over the past decade, deep optical surveys conducted in 50--600 Myr old
open clusters have uncovered numerous young pre-main-sequence stars and
brown dwarfs. New stellar and substellar members have been identified in
clusters, including the Pleiades 
\citep{bouvier98,zapatero99b,dobbie02a,moraux03,bihain06,lodieu07c},
$\alpha$\,Per \citep{stauffer99,barrado02a,lodieu05a}, 
IC\,2391 \citep{barrado01b,barrado04b,spezzi09}, M\,35 \citep{barrado01a}, 
Blanco\,1 \citep{moraux07a}, IC\,4665 \citep{deWit06,cargile10a}, 
the Hyades \citep{dobbie02c,bouvier08a},
and NGC\,2547 \citep{naylor02,jeffries04,jeffries05}.
Cluster membership in these works is generally assessed by proper 
motion \citep{hambly99,adams01a,moraux01,deacon04,lodieu07c}, near-infrared 
photometry \citep{zapatero97b,barrado02a,lodieu07c}, and optical spectroscopy 
for spectral typing, chromospheric activity, and surface gravity 
\citep{steele95a,martin96,lodieu05a,bihain10a}. The $I-J$ and $I-K$ 
optical-to-infrared colours have proven their efficiency to weed out 
background giants and field dwarfs as emphasised by optical-infrared surveys 
carried out in the Pleiades \citep{zapatero97b}, $\alpha$\,Per 
\citep{barrado02a,lodieu05a}, and $\sigma$\,Orionis 
\citep{bejar01,caballero07d}.

The Two Micron All-sky Survey \citep[2MASS;][]{cutri03} and Deep 
Near-Infrared Survey \citep[DENIS;][]{epchtein97} 
provide $JHK$ photometry for all sources down to $K_{s}$\,$\sim$\,14.5 mag, 
enabling us to confirm the membership of selected low-mass cluster 
candidates in numerous regions. However, those panoramic
surveys are too shallow to probe substellar candidates, except in the
nearest and youngest clusters and star-forming regions \citep{tej02}. 

The UKIRT Deep Infrared Sky Survey \citep[UKIDSS;][]{lawrence07}\footnote{More details at www.ukidss.org}, 
made of five sub-surveys, is designed to reach three to four magnitudes 
deeper than 2MASS and cover several thousand of square 
degrees at infrared wavelengths. The UKIDSS project is defined in 
\citet{lawrence07} and uses the Wide Field Camera \citep[WFCAM;][]{casali07} 
installed on the UK InfraRed Telescope (UKIRT) and the Mauna Kea
Observatory \citep[MKO;][]{tokunaga02} photometric system described in 
\citet{hewett06}. The pipeline processing is described in 
Irwin et al.\ (2009, in prep)\footnote{Extensive details on the data
reduction of the UKIDSS images is available at
http://casu.ast.cam.ac.uk/surveys-projects/wfcam/technical, see also
\citet{irwin04}}
and the WFCAM Science Archive (WSA) in \citet{hambly08}.

The Galactic Clusters Survey (GCS), one of the UKIDSS sub-surveys,
aims at probing young brown dwarfs in ten star-forming regions and open
clusters over large areas in five passbands ($ZYJHK$) across the
1.0--2.5 micron wavelength range with a second epoch in $K$.
The main goal is to measure the form of the IMF
\citep{salpeter55,miller79,scalo86} in the substellar regime to tackle 
important issues including the formation and spatial distribution of 
low-mass stars and brown dwarfs. Early results for Upper Scorpius, in 
the Pleiades, $\sigma$ Orionis, and in Praesepe are presented in 
\citet{lodieu06,lodieu07a,lodieu08a,lodieu11a}, \citet{lodieu07c},
\citet{lodieu09e}, and \citep{baker10}, respectively.

Among the open clusters targeted by the GCS is IC\,4665, whose central
coordinates are R.A.\ = 17$^{\rm h}$40 and dec = 05$^{\circ}$. This cluster 
is of special interest because it is among the few pre-main-sequence
open clusters with an age between 10 and 50 Myr. It has suffered little
dynamical evolution and the measured mass function is close to the real IMF
because star-formation processes have ended.
The cluster is young with a lithium age of 27 Myr \citep{manzi08}, consistent 
with the nuclear age estimate \citep[36 Myr;][]{mermilliod81}, activity
probes \citep[$<$40 Myr;][]{montes98}, and other independent studies
\citep[30--40 Myr;][]{lynga95,dambis99,kharchenko95a}.
Several distance estimates are available for IC\,4665. \citet*{abt67a}
found 324 pc from isochrone fitting, \citet{lynga95} reported 430 pc
in his catalogue of open clusters and \citet{mermilliod81} derived
$\sim$350 pc. Independent studies based on Hipparcos observations 
derived a mean distance close to 350 pc: \citet{dambis99} obtained 
370$\pm$100 pc, \citet{hoogerwerf01} reported 385$\pm$40 pc, and 
\citet{kharchenko95a} inferred 352 pc.
In this paper, we will adopt an age of 30 Myr and a distance of 350 pc.
Previous studies of the cluster provided a list of low-mass members based 
on their spectral types, radial velocities, and lithium abundances 
\citep[for a summary, see][]{prosser93}. A recent deep optical ($I$,$z$) 
survey covering the central 3.82 square degrees in the cluster identified 
new candidate members down to 0.03 M$_{\odot}$ \citep{deWit06}. 
Near-infrared photometry was extracted from 2MASS with additional 
photometry obtained for a subsample of brown dwarf candidates with the 
infrared camera on the Canada-France-Hawaii (CFH) telescope.

%
%
\begin{figure}[!h]
   \centering
   \includegraphics[width=0.9\linewidth]{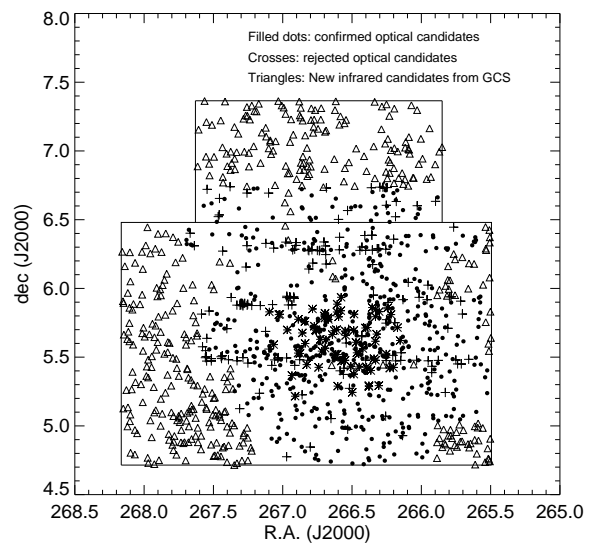}
   \caption{Schematic view of the UKIDSS GCS coverage in
            IC\,4665 is shown with thick solid lines. We included
            known high probability photometric candidates (star symbols)
            from \citet{prosser93}, the optical ($I$,$z$) candidates
            \citet{deWit06} confirmed (filled circles) and rejected
            (plus symbols) by the GCS photometry,
            as well as new photometric candidates identified in the GCS
            outside the optical coverage (open triangles).
            A more detailed view of the optical coverage is shown
            in Fig.\ 2 of \citet{deWit06}.
}
   \label{fig_ic4665_gcs:coverage}
\end{figure}

We present $ZYJHK$ photometry for a $\sim$4.3 square degree
area centred on IC\,4665 obtained with the WFCAM wide-field imager on
UKIRT within the framework of the UKIDSS Galactic Clusters Survey to
investigate the low-mass and substellar mass function.
This paper is organised as follows. In Sect.\ \ref{IC4665_GCS:obs_OPT}
we describe the optical observations obtained with the CFHT in 2002\@. 
In Sect.\ \ref{IC4665_GCS:obs_GCS} we present the GCS survey.
In Sect.\ \ref{IC4665_GCS:corr} we describe the selection of candidates, 
revise the membership of the optically selected candidates on the basis 
of their new infrared colours, and explain the identification of new 
cluster member candidates in the GCS\@. 
In Sect.\ \ref{IC4665_GCS:MF_LF} we derive the luminosity and
mass function for IC\,4665 in the low-mass and substellar regimes.
 
%
%
\section{The optical survey}
\label{IC4665_GCS:obs_OPT}

The optical survey was described in detail in \citet{deWit06}.
We provide here a brief summary to bring our infrared observations
(Sect.\ \ref{IC4665_GCS:obs_GCS}) into context. The open cluster IC\,4665 
was targeted within the framework of a deep, wide-field, photometric survey 
of open clusters and star-forming regions aimed at studying the influence of 
the environment on tthe mass function. The observations, conducted in 
June 2002, surveyed 3.82 square degrees in the cluster down to a completeness 
limit of $I$,$z$ = 22 mag, corresponding to $\sim$0.03 M$_{\odot}$, assuming 
a distance of 350 pc and an age of 50--100 Myr \citep{prosser93}.
We adopt the following notation throughout the paper: $Z$ (in capital letter) 
refers to the WFCAM $Z$ filter with a narrow bandpass (centred at 
0.8 $\mu$m) and $z$ refers to the CFH12K optical $z$ filter. The selection 
of cluster member candidates in the ($I-z$,$I$) colour-magnitude diagram 
yielded a total of 691 low-mass and 94 brown dwarf candidates, assuming the 
age and distance indicated above. The 2MASS catalogue was queried to provide 
infrared photometry for candidates brighter than $K_{s}$ = 14.3 mag. 
Additional infrared ($JHK$) photometry was obtained for a subset of 101 
candidates with the CFHT infrared camera (CFHTIR). The number of
candidates selected from the optical colour-magnitude diagram alone and
later rejected on the basis of their $I-K$ colours is about
35\,\% at faint magnitudes, showing the efficiency of the near-infrared
photometry. The contamination by field stars was also 
estimated from two control fields. The derived mass function after 
correction for statistical contamination was best fit by a lognormal with 
a mean mass of 0.32 M$_{\odot}$, comparable to results in the Pleiades 
\citep{martin98a,dobbie02a,tej02,moraux03,lodieu07c}, and
$\alpha$ Per \citep{barrado02a}.

%
%
\section{The UKIDSS GCS infrared survey}
\label{IC4665_GCS:obs_GCS}

\begin{figure}
   \centering
   \includegraphics[width=\linewidth]{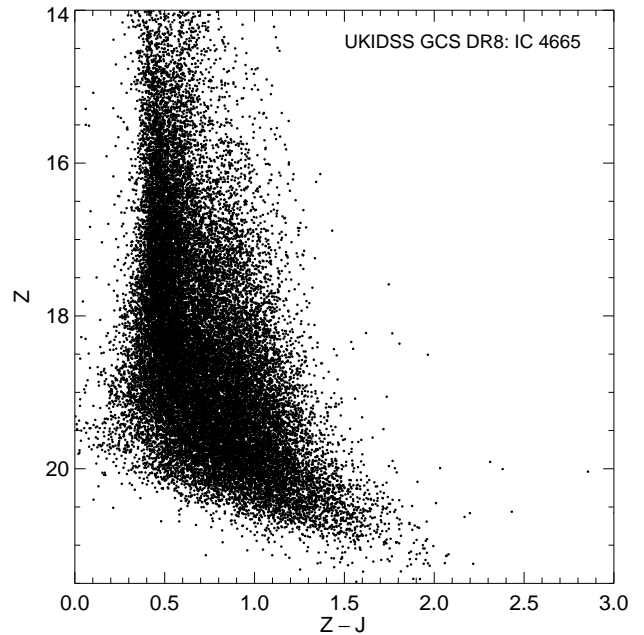}
   \caption{($Z-J$,$Z$) colour-magnitude diagram for 4.3 square degrees 
surveyed in IC\,4665 by the UKIDSS GCS\@. One out of 10 sources is 
plotted.
}
   \label{fig_ic4665_gcs:cmd_zjz_alone}
\end{figure}

We extracted all point sources from the UKIDSS GCS DR8 (September 2010) that are
located within a three degree radius around the centre of IC\,4665 using a 
similar Structure Query Language (SQL) query to our work in the Pleiades 
\citep{lodieu07c}. In summary, we selected only point sources detected
in $ZYJHK$, as well as point sources with $JHK$ photometry, but 
undetected in $Z$ and $Y$. The final catalogue contains 303,978 sources 
distributed in the sky as illustrated in Fig.\ \ref{fig_ic4665_gcs:coverage}. 
The corresponding ($Z-J$,$Z$) colour-magnitude diagram is presented in 
Fig.\ \ref{fig_ic4665_gcs:cmd_zjz_alone} for one out of ten point source. All 
sources fainter than 
$Z$ = 11.5, $J$ = 11.0, and $K$ = 9.9 mag were retrieved but our analysis 
will solely focus on candidates fainter than $J \sim$ 13.3 mag. This limit 
is set by the original selection in the ($I-z$,$z$) colour-magnitude diagram 
\citep[Figure 7 in][]{deWit06} where only objects fainter than $I$ = 15 mag 
were considered, corresponding to a mass of $\sim$0.6 M$_{\odot}$ according 
to 30 Myr isochrones \citep{baraffe98,chabrier00c} shifted to a d = 350 pc.
The GCS survey of IC\,4665 is complete down to $J$ $\sim$ 18.8 mag and 
$K$ $\sim$ 18.1 mag, respectively, adopting a power law fit to the histograms 
of the numbers of point sources as a function of magnitude. We are able to
detect objects a magnitude fainter, which we adopt as our detection limits.

The central coordinates of the tiles observed by the GCS in IC\,4665 are 
listed in Table \ref{tab_ic4665_gcs:log_obs} and the coverage is shown in 
Fig.\ \ref{fig_ic4665_gcs:coverage}. The observations were made in May 2006 
(Table \ref{tab_ic4665_gcs:log_obs}). The coverage of the survey is 
homogeneous over the entire cluster and complete in the $J$ = 12--19 mag
range.

The GCS provides five-band photometry. In addition, we have computed 
proper motions from the $\sim$6 year baseline between the 2MASS 
\citep[observations taken between 26 April and 12 June 2000;][]{cutri03} and 
GCS observations (May 2006) for all sources brighter than $J$ = 15.5 mag. 
Furthermore, we have near-infrared photometry for 313 candidates (only
six have no $Z$-band photomery) out of the 393 optical candidates in the 
$I$=17.0--22.0 magnitude range, i.e.\ 77--78\% of all the faint optical 
candidates whereas \citet{deWit06} presented (only) $K$-band photometry 
for 101 faint candidates (i.e.\ $\sim$25\%).

%
%
%

%
%
\begin{table}
 \centering
  \caption{Log of the observations.}
 \label{tab_ic4665_gcs:log_obs}
 \begin{tabular}{c c c c c c}
 \hline
Tile$^{a}$  &  R.A.$^{b}$  & dec.$^{b}$ & Date \\
 \hline
1 &  17$^{\rm h}$44$^{\rm m}$  &   +06$^{\circ}$02$'$  &  2006-05-10   \\
2 &  17$^{\rm h}$44$^{\rm m}$  &   +05$^{\circ}$10$'$  &  2006-05-10   \\
3 &  17$^{\rm h}$45$^{\rm m}$  &   +06$^{\circ}$55$'$  &  2006-05-11   \\
4 &  17$^{\rm h}$47$^{\rm m}$  &   +06$^{\circ}$02$'$  &  2006-05-10   \\
5 &  17$^{\rm h}$47$^{\rm m}$  &   +05$^{\circ}$10$'$  &  2006-05-10   \\
6 &  17$^{\rm h}$49$^{\rm m}$  &   +06$^{\circ}$56$'$  &  2006-05-11   \\
7 &  17$^{\rm h}$51$^{\rm m}$  &   +06$^{\circ}$02$'$  &  2006-05-11   \\
8 &  17$^{\rm h}$51$^{\rm m}$  &   +05$^{\circ}$10$'$  &  2006-05-10   \\
 \hline
 \end{tabular}
\begin{list}{}{}
\item[$^{a}$] Each WFCAM tile is 54 by 54 arcmin aside and was observed in
all five passbands ($ZYJHK$). Average seeing measured on the images and in
each individual filter ranged from 0.75 to 1.0 arcsec
\item[$^{b}$] Central coordinates (in J2000) of each tile from UKIDSS
GCS DR8
\end{list}
\end{table}
%

%
%
\section{Cross-matching with optical data}
\label{IC4665_GCS:corr}
\subsection{Optical-to-infrared colour-magnitude diagrams}
\label{IC4665_GCS:corr_OPT_IR}

The depth and coverage of the GCS observations allow us to provide infrared 
counterparts in five broad-band filters ($ZYJHK$) for the large majority of 
optical candidates down to the completeness limit of the CFH12K survey 
($\sim$0.03 M$_{\odot}$). As mentioned in the previous section, we have 
photometry for 313 out of the 393 (77--78\%) optical candidates with 
$I$ = 17.0--22.0 mag.

\begin{figure*}
   \centering
   \includegraphics[width=0.49\linewidth]{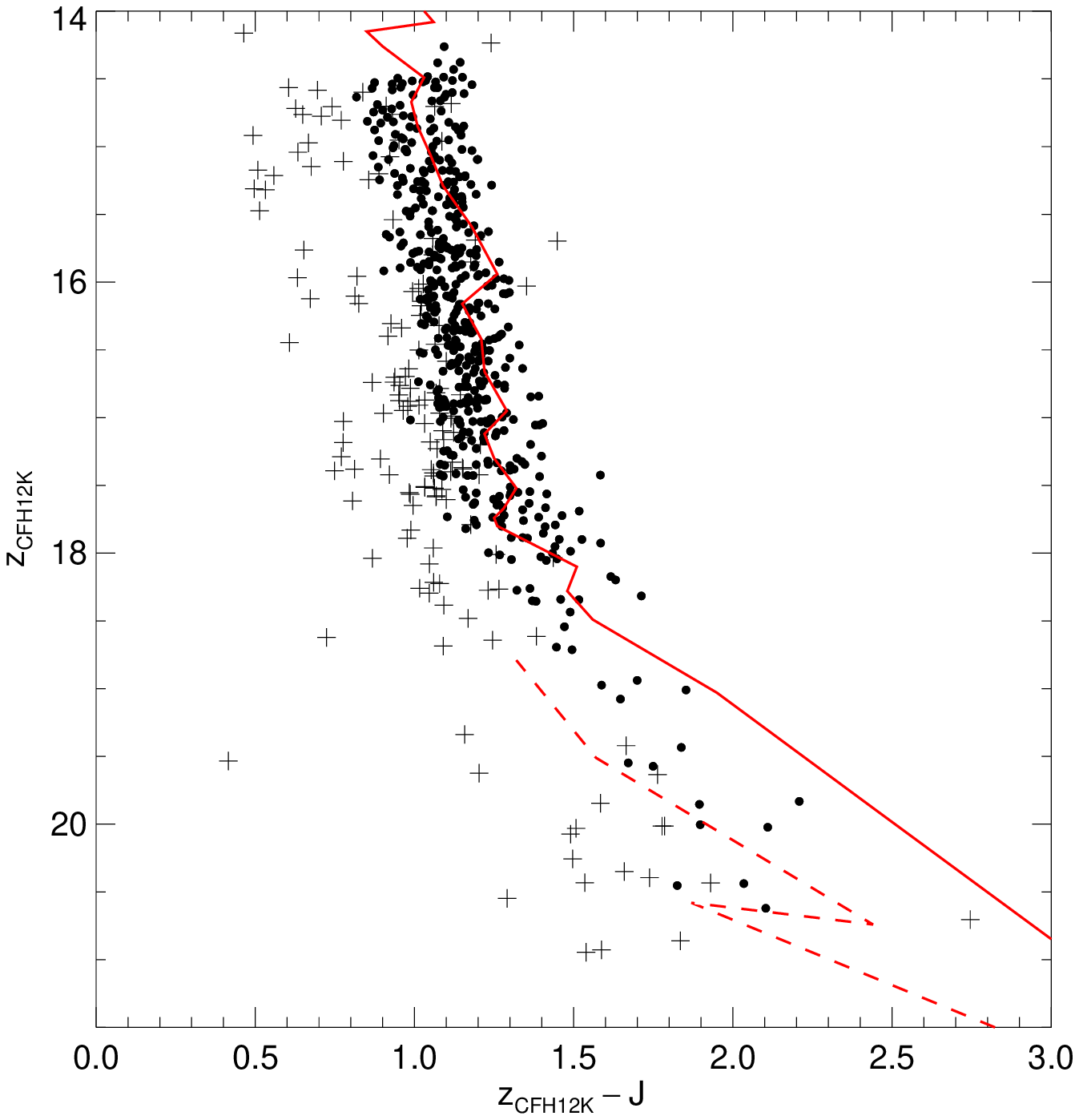}
   \includegraphics[width=0.49\linewidth]{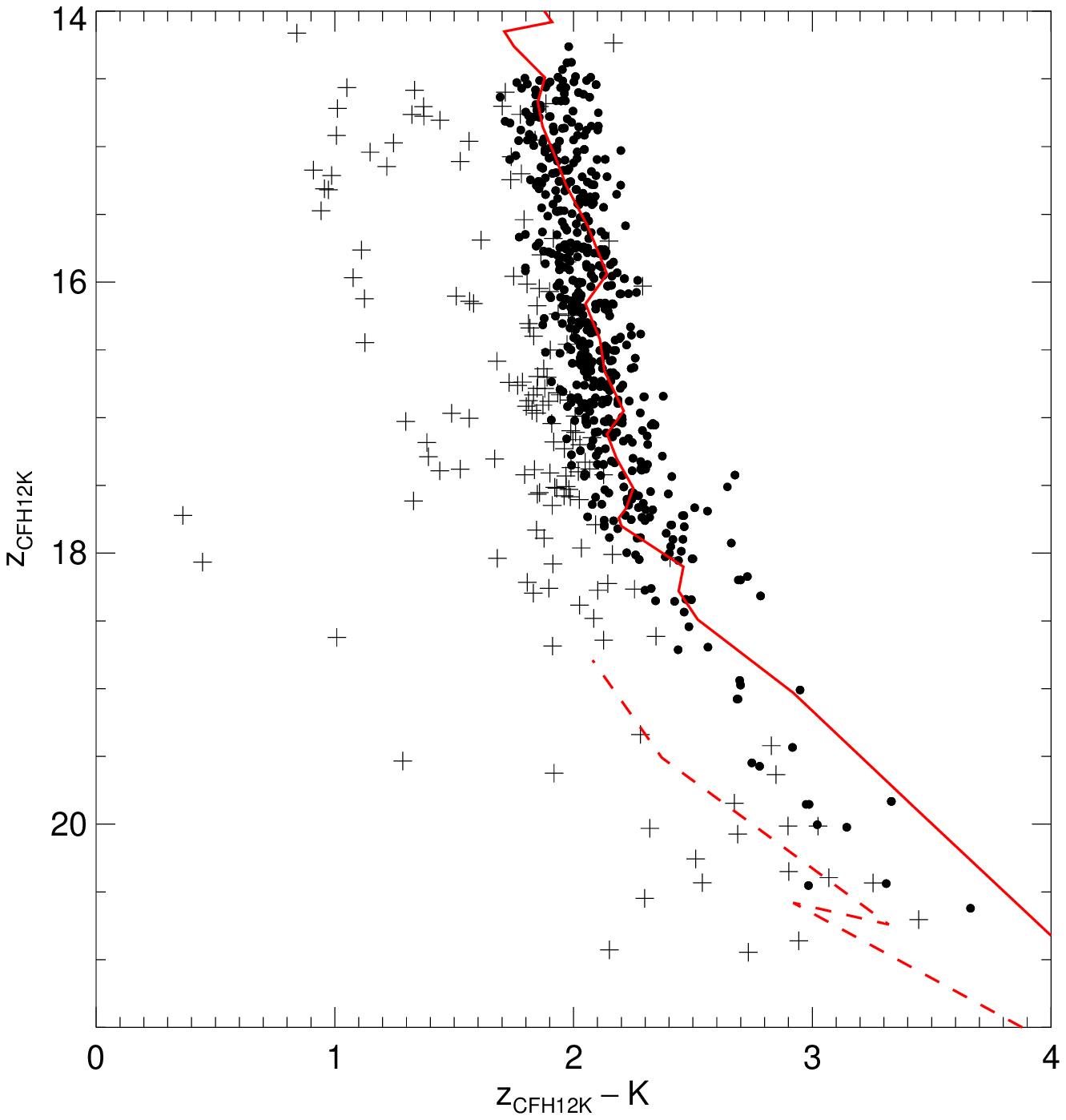}
   \includegraphics[width=0.49\linewidth]{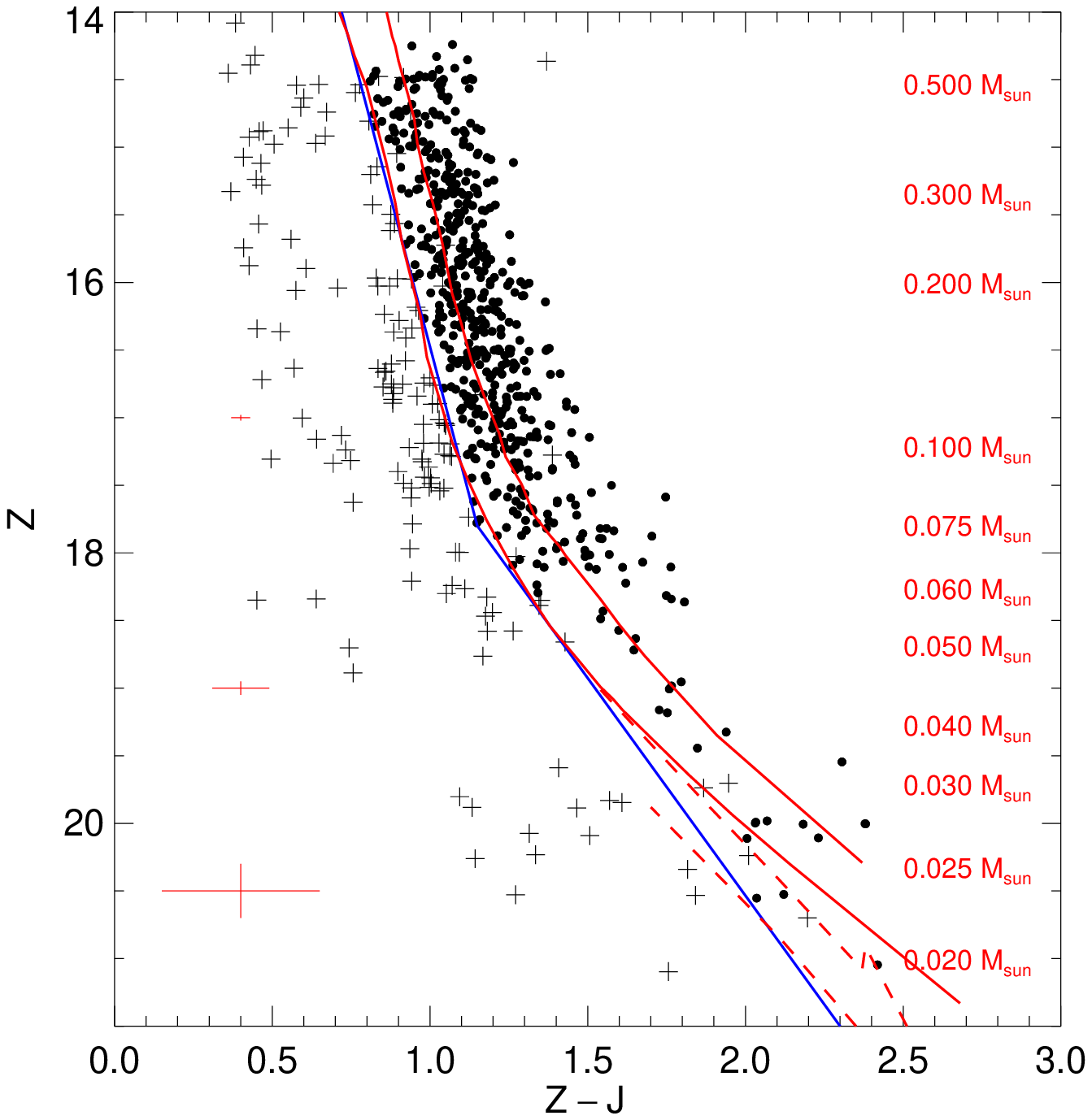}
   \includegraphics[width=0.49\linewidth]{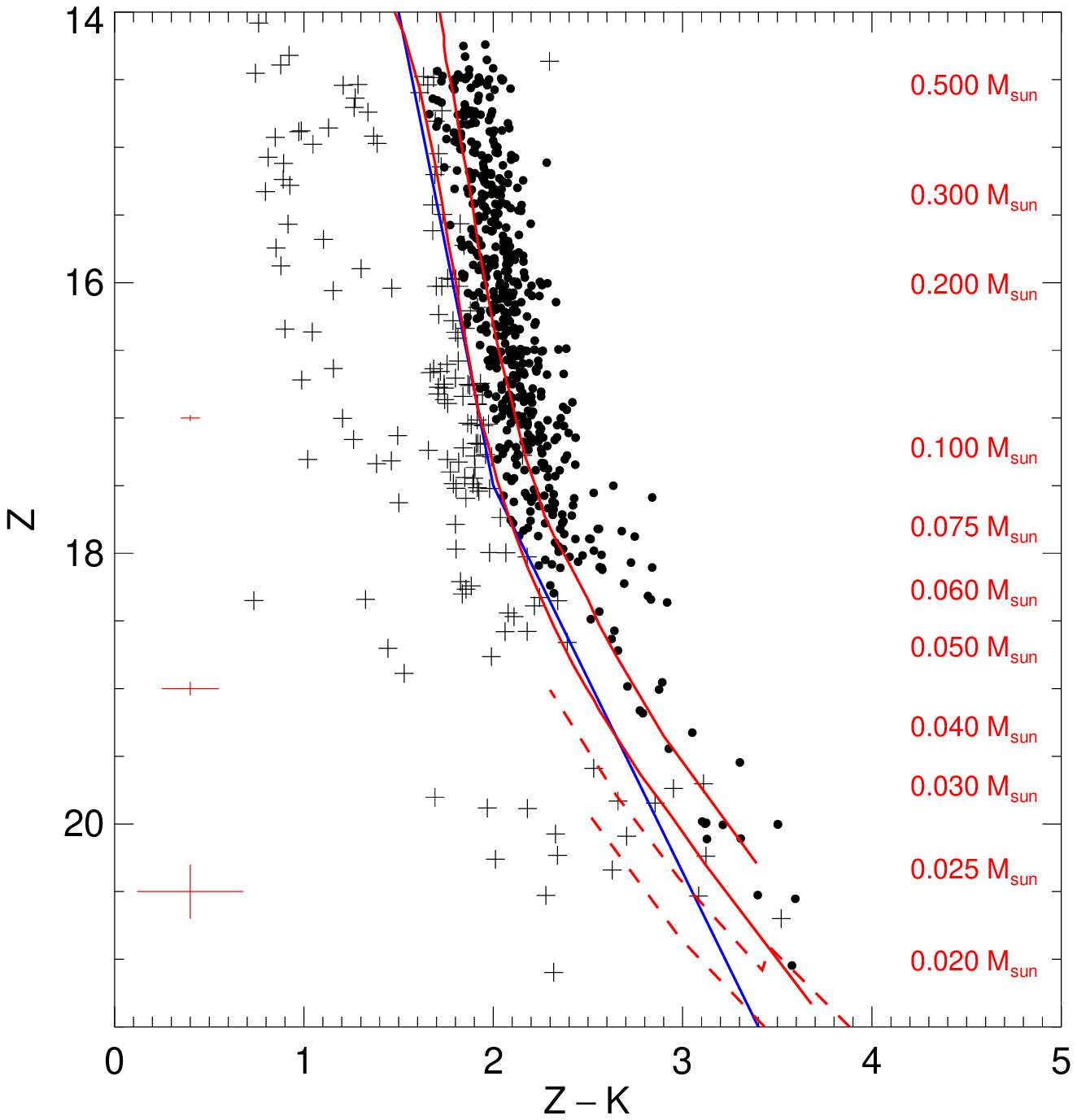}
   \caption{Various colour-magnitude diagrams for $\sim$4.3
square degree surveyed in IC\,4665 by the UKIDSS Galactic Clusters Survey.
Overplotted are the 30 and 100 Myr NextGen \citep[solid line;][]{baraffe98}
and DUSTY \citep[dashed line;][]{chabrier00c} isochrones shifted to a 
distance of 350 pc. Masses are given on the right-hand side of the plot
for the 30 Myr isochrone.
Overplotted are the lines for photometric selection as described in
Section \ref{IC4665_GCS:corr_revised}.
Filled circles represent the optically selected member
candidates from \citet{deWit06} confirmed by infrared photometry.
Crosses are optically selected candidates rejected as cluster
members on the basis of their infrared colours.
{\it{Top left:}} ($z_{CFH12K}-J$,$z_{CFH12K}$) diagram.
{\it{Top right:}} ($z_{CFH12K}-K$,$z_{CFH12K}$) diagram.
{\it{Bottom left:}} ($Z-J$,$Z$) diagram.
{\it{Bottom right:}} ($Z-K$,$Z$) diagram.
Note that the 30 Myr NextGen and DUSTY isochrones
are plotted on the top hand side diagrams with the set
of filters indicated on the axis.
}
   \label{fig_ic4665_gcs:cmd4}
\end{figure*}

We have plotted various colour-magnitude diagrams using optical and infrared 
magnitudes to assess the membership of the optically selected candidates 
published in \citet{deWit06}. On the left-hand side of 
Fig.\ \ref{fig_ic4665_gcs:cmd4} we show the ($z-J$,$z$) and the ($Z-J$,$Z$) 
colour-magnitude diagrams. The same diagrams are displayed for the $z-K$ and 
$Z-K$ colours on the right-hand side of Fig.\ \ref{fig_ic4665_gcs:cmd4}. The 
majority of optical candidates rejected by the $z-J$ and $z-K$ colours are 
also discarded by the $Z-J$ and $Z-K$ colours, showing the efficiency of 
optical-to-infrared colours in removing interlopers 
(Fig.\ \ref{fig_ic4665_gcs:cmd4}).

\begin{figure*}
   \centering
   \includegraphics[width=0.49\linewidth]{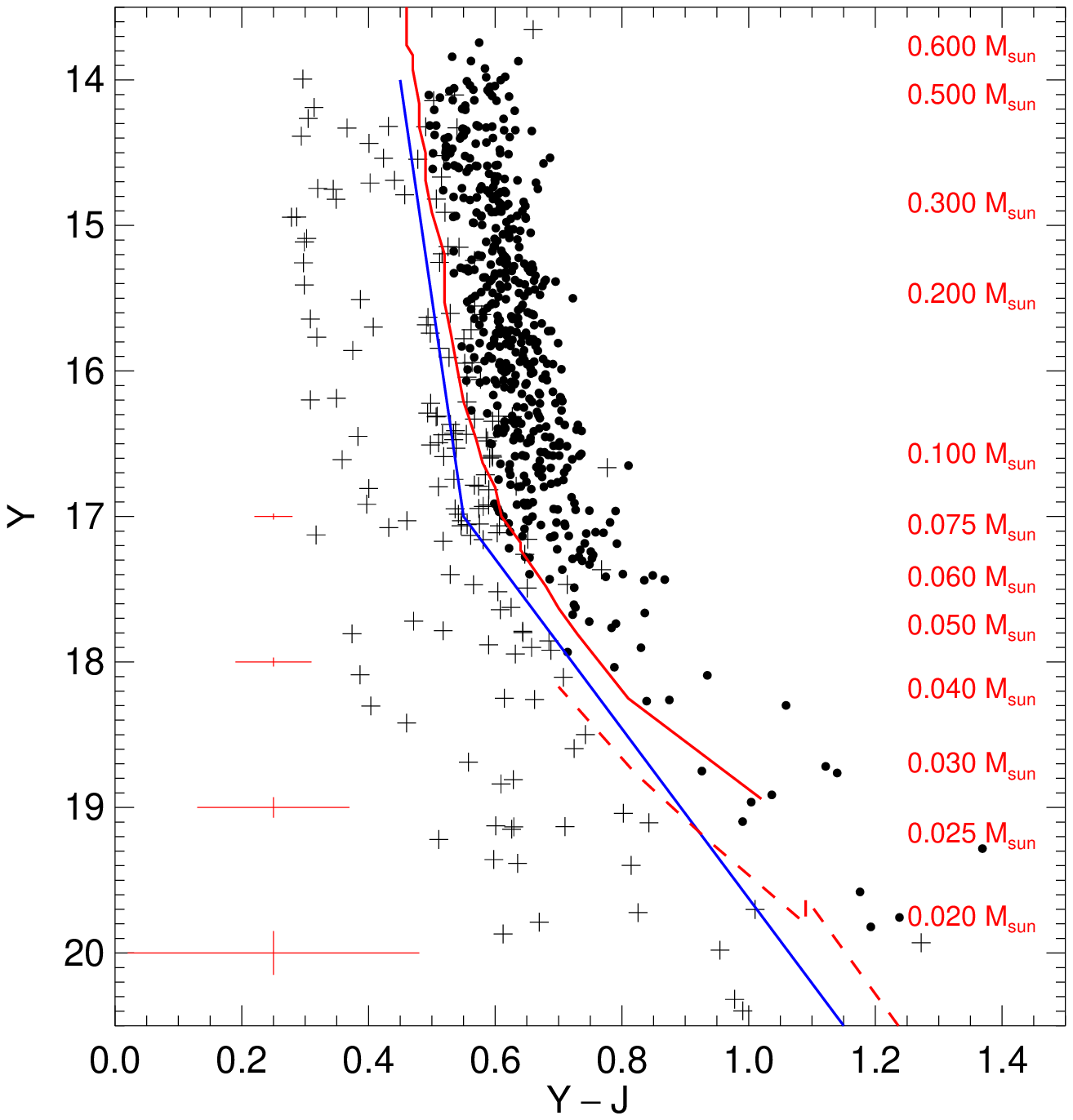}
   \includegraphics[width=0.49\linewidth]{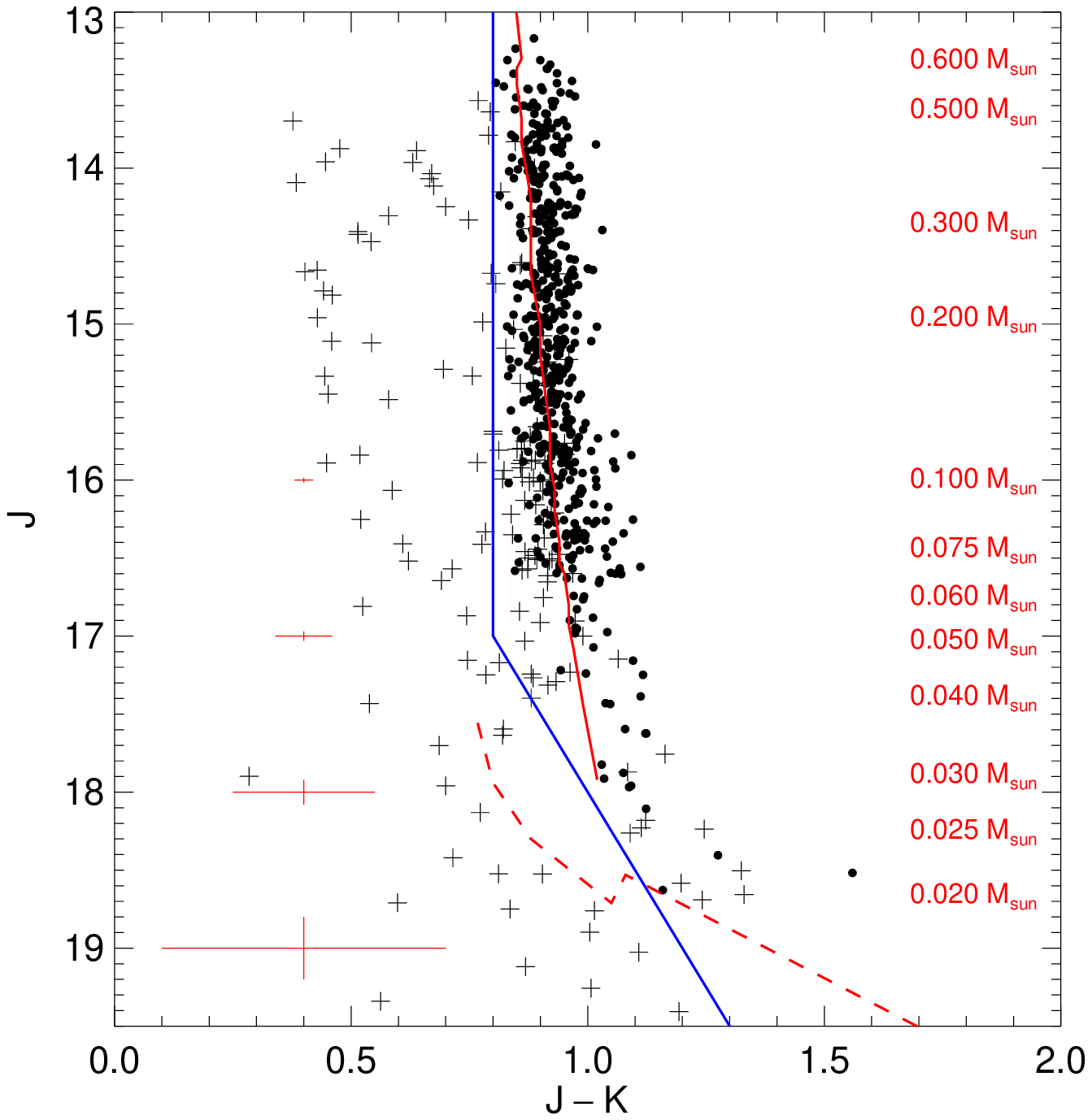}
   \caption{($Y-J$,$Y$) and ($J-K$,$J$) colour-magnitude diagrams 
for a $\sim$4.3 square degree surveyed in IC\,4665 by the GCS\@.
Overplotted are the 30 Myr NextGen \citep[solid line;][]{baraffe98}
and DUSTY \citep[dashed line;][]{chabrier00c} isochrones shifted to a 
distance of 350 pc. Masses are given on the right-hand side of each plot
for the 30 Myr isochrones.
Filled circles represent the optically selected candidates 
from \citet{deWit06} whose membership is confirmed by their infrared 
photometry. Crosses are optically selected candidates rejected as cluster 
members on the basis of their infrared colours.
Overplotted are the lines for photometric selection as described in
Section \ref{IC4665_GCS:corr_revised}.
}
   \label{fig_ic4665_gcs:cmd_yjy_jkj}
\end{figure*}

Additional colour-magnitude diagrams employed to reject photometric
contaminants include the ($Y-J$,$Y$) and ($J-K$,$J$) presented in 
Fig.\ \ref{fig_ic4665_gcs:cmd_yjy_jkj}. We also examined the 
($H-K$,$J-H$) colour-colour diagram to assess the nature of the 
contaminants. This diagram, not shown here, suggests that the contamination 
in the optical selection originates from a mixture of reddened early-type 
dwarfs and field M dwarfs.

%
%
%

%
%
%

%
%
\begin{table}
 \centering
  \caption{Number of optical candidates with GCS photometry (opt)
for different intervals of $Z$ magnitude after the different selection
steps.
}
 \label{tab_ic4665_gcs:contam}
 \begin{tabular}{@{\hspace{0mm}}l c c c c c c c c@{\hspace{0mm}}}
 \hline
Range          & opt   &  S1$^{a}$ &  S2$^{b}$ & S3$^{c}$ &  S4$^{d}$ & S5$^{e}$ & R1$^{f}$ & R2$^{g}$  \cr
 \hline
$Z$ = 14--15   &  91   &  72   &  72   &   72   &  70   &  68   &  74.7  &  94.4 \\
$Z$ = 15--16   & 158   & 139   & 139   &  139   & 138   & 137   &  86.7  &  98.6 \\
$Z$ = 16--17   & 185   & 152   & 152   &  152   & 151   & 150   &  81.1  &  98.7 \\
$Z$ = 17--18   & 139   & 101   & 101   &  101   & 101   & 100   &  71.9  &  99.0 \\
$Z$ = 18--19   &  42   &  28   &  25   &   24   &  24   &  24   &  57.1  &  85.7 \\
$Z$ = 19--20   &  15   &   9   &   9   &    7   &   7   &   7   &  46.7  &  77.7 \\
$Z$ = 20--21   &  15   &   8   &   8   &    6   &   6   &   6   &  40.0  &  75.0 \\
 \hline
$Z \leq$ 17.2  & 479   & 390   & 390   &  390   & 386   & 382   &  79.7  &  97.9 \\
$Z \geq$ 17.2  & 175   & 120   & 117   &  112   & 112   & 111   &  63.4  &  92.5 \\
 \hline
$Z$ = 14--21   & 654   & 510   & 507   &  502   & 498   & 493   &  75.4  &  96.6 \\
 \hline
 \end{tabular}
\begin{list}{}{}
\item[$^{a}$] Number of optically selected candidates kept after
applying the $Z-J$ selection (step 1; S1)
\item[$^{b}$] Number of candidates kept after the $Z-K$ selection (step 2; S2)
\item[$^{c}$] Number of candidates kept after the $Y-J$ selection (step 3; S3)
\item[$^{d}$] Number of candidates kept after the $Y-K$ selection (step 4; S4)
\item[$^{e}$] Number of candidates kept after the $J-K$ selection (step 5; S5)
\item[$^{f}$] The R1 percentage refers to the S5/opt ratio
\item[$^{g}$] The R2 percentages refers to the S5/S1 ratio
\end{list}
\end{table}
\subsection{Revised membership of optically selected candidates}
\label{IC4665_GCS:corr_revised}

To verify the membership of the 785 optical candidates from \citet{deWit06} 
using our near-infrared photometry, we cross-matched that list with 
the full GCS catalogue with a matching radius of five arcsec and found 654 
common objects. We found that 58 out 785 optical candidates lie 
in the region outside of GCS coverage. Therefore, 785$-$58=727 objects are 
within the GCS coverage, implying that 727$-$654=73 sources are not 
recovered by our GCS selection. The GCS detection of these 73 sources, 
located at the position of the optical source, did not satisfy the criteria 
set in the SQL query. We found that 19 of the 73 objects are missing 
photometry in (at least) one of the $JHK$ passbands, and the remaining 
sources do not satisfy the point sources criteria ({\tt{Class}} and 
{\tt{ClassStat}} parameters in at least one filter). This problem is inherent 
to the SQL query and represents about 10\% of the total number of candidates, 
a value slightly higher than for \sOri{} \citep[4.4--7.8\%;][]{lodieu09e}. 
We note that 35 out of the 54 ($\sim$65\%) sources with $ZYJHK$ magnitudes 
from GCS DR8 would remain as photometric member candidates after applying 
the colour selections described in the next paragraph.

We generated various colour-magnitude diagrams
(Figs.\ \ref{fig_ic4665_gcs:cmd4} \& \ref{fig_ic4665_gcs:cmd_yjy_jkj})
to design selection cuts based on the position of the 654 optical candidates 
with $ZYJHK$ photometry. Looking at these candidates in the several 
colour-magnitude diagrams, we generated a subsample of candidates by 
extracting only those to the right of lines defined in each diagram below:
\begin{itemize}
\item[$\bullet$] ($Z-J$,$Z$) = (0.72,14.0) to (1.15,17.8)
\item[$\bullet$] ($Z-J$,$Z$) = (1.15,17.8) to (2.3,21.5)
\item[$\bullet$] ($Z-K$,$Z$) = (1.5,14.0) to (2.0,17.5)
\item[$\bullet$] ($Z-K$,$Z$) = (2.0,17.5) to (3.4,21.5)
\item[$\bullet$] ($Y-J$,$Y$) = (0.45,14.0) to (0.55,17.0)
\item[$\bullet$] ($Y-J$,$Y$) = (0.55,17.0) to (1.15,20.5)
\item[$\bullet$] ($Y-K$,$Y$) = (1.3,14.0) to (1.45,17.0)
\item[$\bullet$] ($Y-K$,$Y$) = (1.45,17.0) to (2.2,20.5)
\item[$\bullet$] ($J-K$,$J$) $\geq$ 0.8 for $J$ = 13--17 mag
\item[$\bullet$] ($J-K$,$J$) = (0.8,17.0) to (1.3,19.5)
\end{itemize}

We are left with 493 optical candidates that satisfy those criteria, and
accordingly we confirm their candidacy as photometric members on the basis of 
the five-band infrared photometry. This sample will be refered to as 
``high-probability members'' in the rest of the paper. These members have a 
magnitude range that spans $Z$ = 14.2--21.1 mag. Their coordinates, photometry 
and proper motions are available electronically and an example is shown in 
Table \ref{tab_ic4665_gcs:NIR_opt_cand}. They are plotted as filled circles 
in Figs.\ \ref{fig_ic4665_gcs:cmd4} and \ref{fig_ic4665_gcs:cmd_yjy_jkj}.
The other optical candidates, rejected as cluster members, are listed
in Table \ref{tab_ic4665_gcs:NIR_opt_NM} and plotted as plus symbols
in Figs.\ \ref{fig_ic4665_gcs:cmd4} \& \ref{fig_ic4665_gcs:cmd_yjy_jkj}.
Tables \ref{tab_ic4665_gcs:NIR_opt_cand} and \ref{tab_ic4665_gcs:NIR_opt_NM}
list the names of each source according to the International Astronomical 
Union (IAU), the coordinates (in J2000), the optical ($Iz$) and infrared
($ZYJHK$) magnitudes, as well as the proper motions in right ascension
and declination expressed in mas/yr. We emphasise that proper motions
for objects fainter than $J$ = 15.5 mag are unreliable because they are
computed using 2MASS as first epoch and the GCS as a second epoch.

We compared the percentage of contaminants as a function of magnitude 
before and after applying the colour cuts above: we observe an increase in 
the level of contamination (from $\sim$20\% to $\sim$50\%) with decreasing 
luminosity in particular below $Z$ = 17 mag (corresponding to 
0.15 M$_{\odot}$; Table \ref{tab_ic4665_gcs:contam}), in agreement with the 
conclusions drawn by \citet{deWit06} from a limited sample of $K$-band 
observations and the study of two control fields \citep[see also][]{moraux03}
in this open cluster located at lower galactic latitude and towards a dense
region \citep[see e.g.][]{prosser93}. The most powerful selection criterion 
clearly is the $Z-J$ colour because the other colour cuts remove on average less 
than 10\% of the candidates (R2 percentage in Table \ref{tab_ic4665_gcs:contam}).

We looked further into the level of contamination expected in the GCS 
data. We selected two control fields with galactic coordinates similar 
to IC\,4665\@. These are located about 0.2 degrees outside the
cluster radius determined by \citet{kharchenko05a}. The two control
fields contain 6610 and 7725 sources with $ZYJHK$ photometry, respectively,
compared to a total of 7220 sources in the central 0.16 square degrees
in IC\,4665\@. After applying the various selection cuts designed above, 
we are left with 89 candidates in the central region of the cluster compared
with 57 and 50 candidates in the two control fields. The average
level of contamination is of the order of 55--60\% and is comparable
in the bright and faint magnitude ranges. This level of contamination
is very likely an upper limit because the control fields are located very
close to the cluster, much closer than the control fields used in the
optical study (0.2 vs 3 degrees). The estimated contamination is higher
than the levels derived for the Pleiades \citep{moraux01} and Alpha
Per \citep{barrado02a}, most likely because of the lower galactic latitude
of IC\,4665\@.

%
%
\subsection{New cluster member candidates}
\label{IC4665_GCS:corr_new_Members}

Considering the low-mass and substellar sequence defined by the
high-probability members, we have identified new candidates over the full 
area covered by the GCS\@. The same colour cuts as defined in 
Sect.\ \ref{IC4665_GCS:corr_revised} returned a total of 2236 sources with 
$Z$ = 14.0--21.25 mag. Of the 2236 sources, 864 are outside the optical 
coverage whereas 1372 are located in the common area between the optical and 
GCS surveys, including the 493 optical candidates confirmed as photometric 
members by the GCS (Sect.\ \ref{IC4665_GCS:corr_revised}). Hence, we have 
1372$-$493=879 potential new candidates (643 are brighter than $Z$ = 17.2 
mag and 236 fainter). 

What about these new objects? Many lie in the blue 
part of the selection in the ($Z-J$,$Z$) diagram, suggesting that the large 
majority are indeed non-members. This trend is also observed in the 
($Y-J$,$Y$) diagram and clearly confirmed in the optical 
($i-z_{\rm CFHT}$,$i_{\rm CFHT}$) diagram. Furthermore, 635 and 30 (=665) 
out of 879 have poor quality flags in 
the $i$ and $z$ filters, respectively. Of these 214 candidates, only 26
satisfy the colour criteria set by \citet{deWit06}: 23 were not selected 
by the optical study because their quality flags were greater than 1,
another one was excluded from the criteria set for the flux radius
parameter in SEXtractor, and the remaining two (UGCS J174823.92$+$055254.3
and 174543.44$+$062225.5) were overlooked by \citet{deWit06}. Therefore, 
if we consider all 879 (or 879$-$2) candidates selected from the GCS but 
unselected by the CFH12K survey, the level of contamination of the GCS 
selected in IC\,4665 would be of the order of 64\%(879/1372 or 877/1372),
consistent with the 55--60\% estimated from the control fields 
((Sect.\ \ref{IC4665_GCS:corr_revised}).

Furthermore, we have 2236$-$1372=864 new candidates from the GCS
located outside the optical coverage, including 664 brighter than
$Z$ = 17.2 mag and 200 fainter (Table \ref{tab_ic4665_gcs:newMEMB}). 
If we consider the level of contamination inferred above (55--64\%),
we should expect around 239--299 probable members among 664 bright 
candidates and about 72--90 out of 200 at fainter magnitudes, but these 
estimates are purely statistical. Nevertheless, we can say that the 
probability of membership decreases with increasing distance from the 
selection line.

%
%
\subsection{Radial distributions}
\label{IC4665_GCS:corr_radial_distribution}

\begin{figure*}
   \centering
   \includegraphics[angle=270, width=0.49\linewidth]{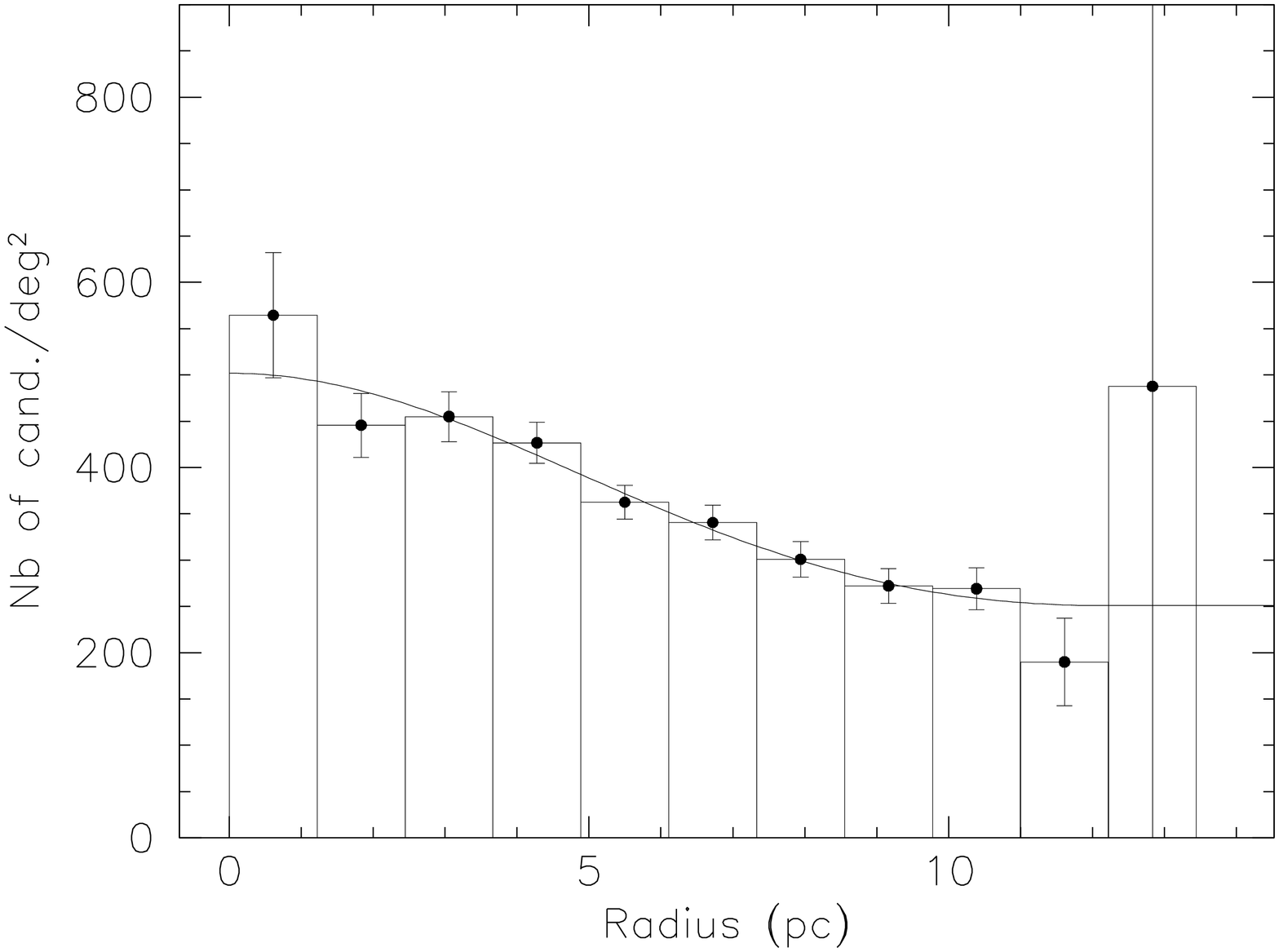}
   \includegraphics[angle=270, width=0.49\linewidth]{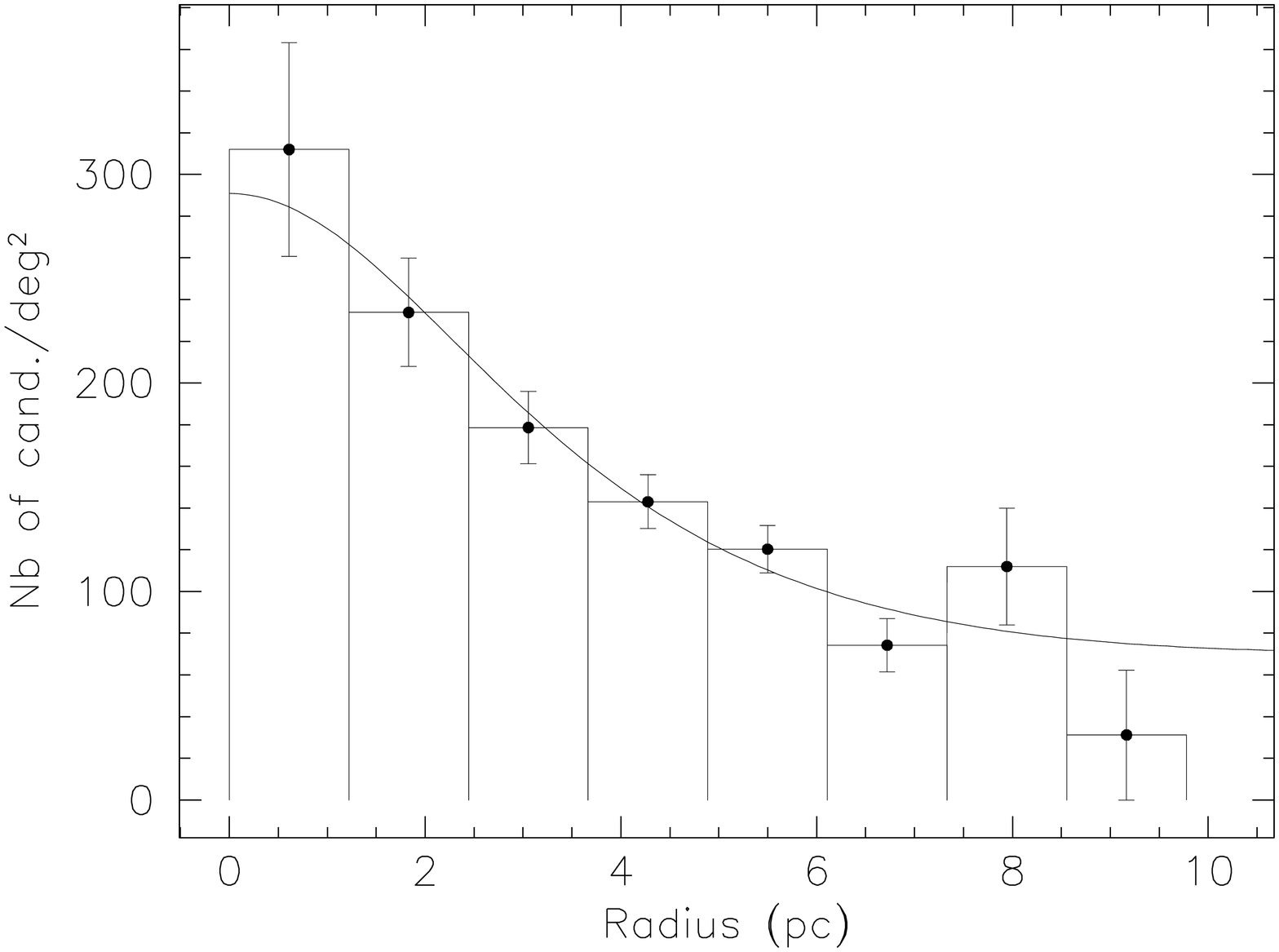}
   \caption{Radial distributions for the full GCS coverage (left) and the
common area between the GCS and the CFH12K surveys (right).
}
   \label{fig_ic4665_gcs:Radial_distrib}
\end{figure*}

To estimate the spatial distribution of the candidates selected from the 
UKIDSS data alone, we calculate the number of objects per square degree 
located within an annulus $R, R+dR$ centred on the cluster centre 
$\alpha_0=17^h46^m04^s, \delta_0=5^d38^m53^s$ \citep{deWit06}. The
corresponding radial profile is shown in the left-hand side of
Figure \ref{fig_ic4665_gcs:Radial_distrib}. This
distribution is then fitted by a King profile \citep{king62} plus a
constant assuming that the density of contaminants $n_{cont}$ is uniform:
\begin{equation}
n(x) = k\,\left[\frac{1}{\sqrt{1+x}} - \frac{1}{\sqrt{1+x_{t}}}\right]^2 + n_{cont},
\end{equation} 

where $k$ is a normalisation constant, $x=(r/r_{c})^2$ and
$x_{t}=(r_{t}/r_{c})^2$. The distance from the cluster centre is $r$,
$r_{c}$ the core radius and $r_{t}$ the tidal radius.

To find the best fit minimizing $\chi^2$, we left all these parameters free, 
except for the tidal radius that we forced to be between 6.4 and 12.1 pc. 
Indeed, \citet{piskunov08a} found the cluster tidal radius to be 
$r_t=8.7\pm2.3$ pc or $10.7\pm1.4$ pc whether they measured it or calculated 
it from the semi-major axis. We thus found $k$ = 2924$\pm$189 sq.deg.$^{-1}$, 
$r_c$ = 12.1$\pm0.4$ pc, $r_t$ = 12.1$\pm0.3$ pc and $n_{cont}= 251\pm8$ 
per square degree. The total covered area being 6.26 square degrees, this 
corresponds to 1571$\pm$50 contaminants over 2236 candidates, or a 
contamination level of 70.3$\pm$2.3\%, which is consistent with our 
estimate using control fields.

We performed the same analysis for the sample of candidates common to both 
the GCS and the CFH12K survey. The best fit gives $k$ = 497$\pm$40 
sq.deg.$^{-1}$, $r_c$ =4.3$\pm0.8$ pc, $r_t$ = 12.1$\pm0.8$ pc and
$n_{cont}$ = 71$\pm$7 per square degree (see right panel in 
Figure \ref{fig_ic4665_gcs:Radial_distrib}). 
Again using that the common area is 3.475 square degrees, 
we find a contamination level of 50$\pm$5\%.

%
%
\subsection{Proper motions for the brightest member candidates}
\label{IC4665_GCS:corr_PM}

The mean proper motion of the IC\,4665 cluster is estimated to be
($\mu_{\alpha}\cos{\delta}$,$\mu_{\delta}$)=($-$0.57$\pm$0.30,$-$7.40$\pm$0.36)
mas/yr from 30 members included in the study of \citet{kharchenko05a}, in 
agreement with the average proper motion of 13 members measured by Hipparcos 
\citep{hoogerwerf01}. For more details, we refer the reader to the discussion
in Section 4.1 of \citet{deWit06}. Hence, the cluster's motion is low and
not easy to separate from field stars and reddened background giants.
However, astrometric information can be used to reject photometric
candidates with large proper motions.

\begin{figure}
   \centering
   \includegraphics[width=\linewidth]{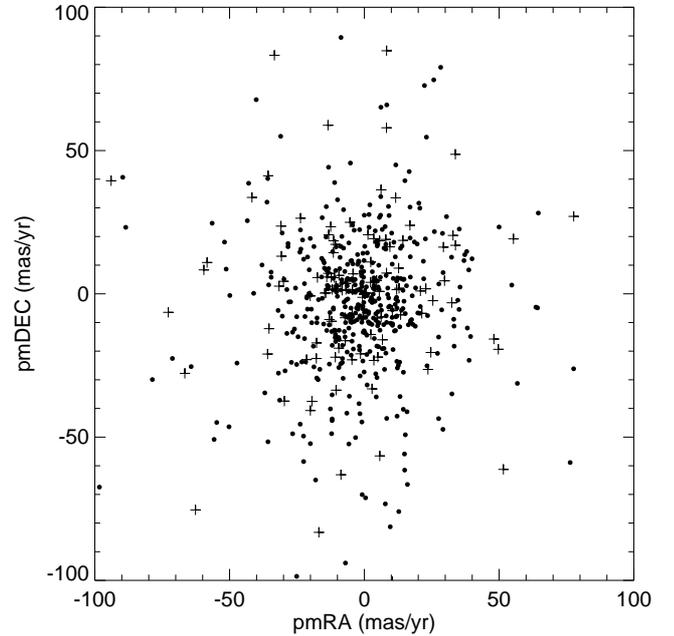}
   \caption{Vector point diagram for all optical candidates confirmed
as probable members by the GCS photometry (black dots) and photometric
non-members (crosses). Typical error bars on the proper motion are 15 mas/yr.
}
   \label{fig_ic4665_gcs:VPDplot}
\end{figure}

We derived proper motions for all point sources identified in IC\,4665 
and brighter than $J$ = 15.5 mag using 2MASS and the GCS as first and second 
epoch, respectively, thanks to the $\sim$6 year baseline. We estimated
the root mean square of the cross-match to better than 15 mas/yr for 
$J$ $\leq$ 15.5 mag. We selected potential proper motion members among
the photometric candidates, adopting a 3$\sigma$ (corresponding to a total
proper motion less than 45 mas/yr; Fig.\ \ref{fig_ic4665_gcs:VPDplot})
cut, which should optimize the selection \citep{lodieu11a}. 

Assuming those parameters, we are left with
278 out the 316 optical candidates brighter than $J$ = 15.5 mag confirmed by 
the GCS as members, yielding a 10\% contamination. Repeating this process for
the new GCS candidates lying outside the CFH12K area, we would select
460 out of the 501 candidates with $J$ $\leq$ 15.5 mag, again suggesting
a contamination of 10\%. A similar result is obtained for the sample of
the GCS candidates with optical photometry that were not selected as potential
members by the optical survey. To summarise, the overall level of contamination
by proper motion non-members is of the order of 10\%.

%
%
\subsection{Fainter cluster member candidates}
\label{IC4665_GCS:corr_fainter_Members}

We attempted to identify fainter and lower mass candidates using sources
that are $Z$-band non-detections. We applied the same colour criteria in
$Y-J$, $Y-K$, and $J-K$ colours as presented in 
Sect.\ \ref{IC4665_GCS:corr_revised}. However, we limited our selection
to $J$ = 18.8 mag, the completeness of the GCS towards IC\,4665\@. At this
magnitude, point sources have a signal-to-noise ratio of the order of 6--8\@.
The SQL query returned three potential candidates after removing those
clearly detected on the $Z$ images. However, two of them are actually detected
on the deep CFH12K images (Table \ref{tab_ic4665_gcs:faint_cand}), suggesting 
that they do not belong to the cluster. These three candidates are listed at 
the top of Table \ref{tab_ic4665_gcs:faint_cand}.

We repeated the same procedure for $Z$ and $Y$ non-detections, using the
$J-K$ colour as sole criterion. The query returned two good candidates after
removal of objects located at the edge of the detector and those clearly
detected on the $Z$ and $Y$ images. These two candidates are listed at
the bottom of Table \ref{tab_ic4665_gcs:faint_cand}.

We repeated those selections in the cluster centre and the two control fields,
but the few objects returned in each area were all detected in $Z$ for the 
$YJHK$ candidates and in $Z$ and $Y$ for the $JHK$ candidates.

%
%
%
%
\begin{table}
 \centering
  \caption{Faint $YJHK$ (top) and $JHK$ (bottom) candidates identified
in the UKIDSS GCS\@.
}
 \label{tab_ic4665_gcs:faint_cand}
 \begin{tabular}{c c c c c c}
 \hline
R.A.$^{a}$ &  dec$^{a}$ &  $Y^{a}$ &  $J^{a}$ &  $H^{a}$ & $K^{a}$ \cr
 \hline
17:43:18.53 & +05:14:34.9 & 19.929 & 18.657 & 18.047 & 17.326$^{b}$ \\
17:46:27.11 & +06:18:08.8 & 19.919 & 18.615 & 18.187 & 17.269$^{b}$ \\
17:50:28.78 & +05:51:16.9 & 19.379 & 18.265 & 17.436 & 16.820$^{c}$ \\
 \hline
17:42:04.01 & +06:15:56.6 &  ---   & 18.783 & 18.023 & 16.745$^{c}$ \\
17:48:42.73 & +05:00:57.6 &  ---   & 18.607 & 17.855 & 17.132$^{c}$ \\
 \hline
 \end{tabular}
\begin{list}{}{}
\item[$^{a}$] Coordinates and photometry from UKIDSS GCS DR8
\item[$^{b}$] Detected on the deep CFH12K $i$ and $z$ images
\item[$^{c}$] Undetected on the deep CFH12K images
\end{list}
\end{table}
%

%
%
\section{The luminosity and mass functions}
\label{IC4665_GCS:MF_LF}

In this section, we derive the cluster luminosity and mass functions based 
on the sample of optical photometric candidates confirmed by the GCS 
photometry. This sample contains the 493 cluster member candidates
discussed in Sect.\ \ref{IC4665_GCS:corr_revised}. We did not correct
the luminosity and mass function for the level of contamination estimated
in section \ref{IC4665_GCS:corr_PM} because this sample represents the 
highest quality sample extracted to date. We expect a low level
of contamination in this photometric sample, which combines optical and
near-infrared photometry: about 10\% of them may be proper motion non-members 
(see Section \ref{IC4665_GCS:corr_PM}) and another 5--10\% spectroscopic
non-members according to our spectroscopic study of Upper Sco \citep{lodieu11a}.
An independent photometric and spectroscopic study of Blanco\,1 by 
\citet{moraux07a} suggests that the expected number of contaminants is low 
after combining optical and infrared photometric bands.

\subsection{The luminosity function}
\label{IC4665_GCS:LF}

This sample of optical photometric candidates with infrared photometry
contains 493 sources in the $J$ $\sim$ 13.1--18.6 mag. Adopting the lithium 
age of 27 Myr derived by \citet{manzi08} for IC\,4665, the stellar/substellar 
limit is at $K$ = 15.55 mag (or $J$ = 16.49 mag) adopting a distance of 
350 pc and choosing the 30 Myr available in the NextGen 
\citep{baraffe98} and DUSTY \citep{chabrier00c} models to be most consistent
with the age of the cluster. Thus, our sample contains $\sim$91\% of stars 
and 8--9\% (40--46 out of 493) substellar candidates.

\begin{figure}
   \centering
   \includegraphics[width=\linewidth]{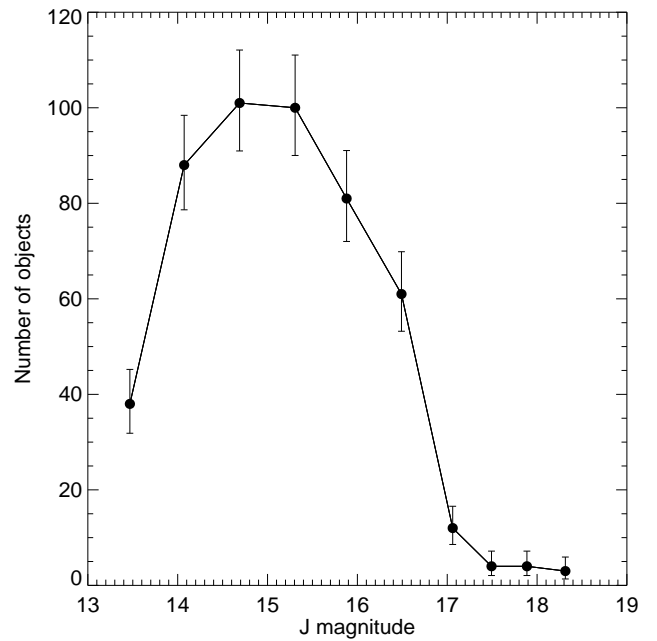}
   \caption{Cluster luminosity function: number of objects
as a function of $J$ magnitude (Table \ref{tab_ic4665_gcs:LF_MF_numbers}). 
Error bars are Gehrels errors.
}
   \label{fig_ic4665_gcs:LFplot}
\end{figure}

Figure \ref{fig_ic4665_gcs:LFplot} displays the luminosity function, i.e.\ the 
number of stars as a function of $J$ magnitude. The number of objects 
increases up to $J \sim$ 15 mag, then decreases down to $J \sim$ 17.5 mag,
and remains flat afterwards. Error bars
are Gehrels errors \citep{gehrels86} rather than Poissonian error
bars because the former represent a better approximation to the
true error for small numbers. The upper limit is defined as
1+($\sqrt(dN+0.75)$) and the lower limit as $\sqrt(dN-0.25)$
assuming a confidence level of one sigma.

%
%
%

%
%
\begin{table}
 \centering
  \caption{Numbers for the luminosity and mass functions.}
 \label{tab_ic4665_gcs:LF_MF_numbers}
 \begin{tabular}{@{\hspace{0mm}}c c c c c c@{\hspace{0mm}}}
 \hline
$\Delta$Mag$^{a}$ & $J_{c}^{b}$ & $\Delta$Mass$^{c}$ & dN$^{d}$ & dN/dM$^{e}$  & dN/d$\log$M$^{e}$ \cr
 \hline
13.160--13.776  &  13.468  &  0.660--0.470   &  38  &  200.4 &   257.7   \\
13.776--14.368  &  14.072  &  0.470--0.310   &  88  &  550.0 &   486.9   \\
14.368--15.010  &  14.689  &  0.310--0.200   & 101  &  918.2 &   530.7   \\
15.010--15.605  &  15.308  &  0.200--0.135   & 100  & 1538.5 &   585.8   \\
15.605--16.155  &  15.880  &  0.135--0.093   &  80  & 1904.8 &   494.3   \\
16.155--16.826  &  16.491  &  0.093--0.059   &  62  & 1550.0 &   253.9   \\
16.826--17.296  &  17.061  &  0.059--0.044   &  12  & 1333.3 &   148.5   \\
17.296--17.690  &  17.493  &  0.044--0.035   &   4  &  444.4 &    40.2   \\
17.690--18.084  &  17.887  &  0.035--0.028   &   4  &  571.4 &    41.3   \\
18.084--18.546  &  18.315  &  0.028--0.022   &   3  &  500.0 &    28.6   \\
 \hline
 \end{tabular}
\begin{list}{}{}
\item[$^{a}$] Magnitude ranges in the $J$-band
\item[$^{b}$] Central magnitudes ($J_{c}$)
\item[$^{c}$] Mass ranges ($\Delta$Mass). The mass intervals were chosen 
identical to those in \citet{deWit06} for direct comparison.
We assumed an age of 30 Myr and a distance of 350 pc for IC\,4665
\item[$^{d}$] Numbers of objects per magnitude interval
\item[$^{e}$] Mass functions in linear (dN/dM) and logarithmic 
(dN/d$\log$M) scales
\end{list}
\end{table}

%
%
%
\begin{figure}
   \centering
   \includegraphics[width=\linewidth]{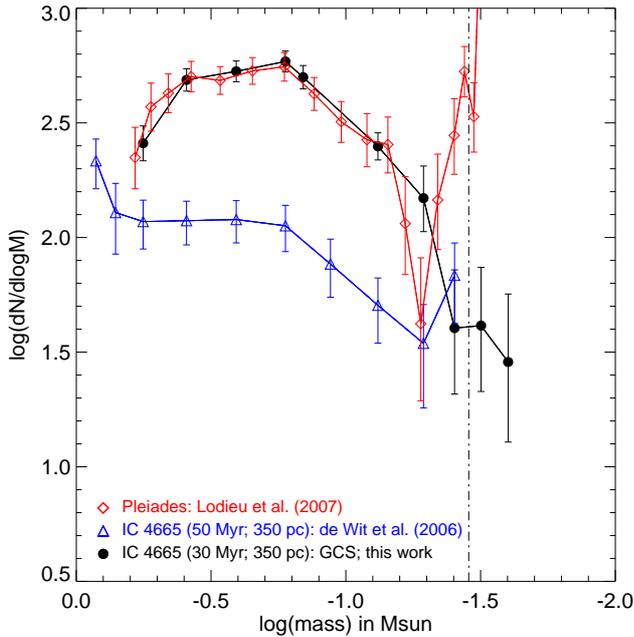}
   \caption{Cluster mass function assuming an age of
30 Myr \citep{manzi08} and a distance of 350 pc.
The conversion from magnitudes
into masses made use of the NextGen and DUSTY models
\citep{baraffe98,chabrier00c}. Values are given in
Table \ref{tab_ic4665_gcs:LF_MF_numbers}.
The vertical dot-dashed lines represent the GCS completeness
limits in IC\,4665\@.
}
   \label{fig_ic4665_gcs:MFplot}
\end{figure}
\subsection{The mass function}
\label{IC4665_GCS:MF}

To convert magnitudes into masses we used the NextGen and DUSTY
models at 30 Myr from the Lyon group \citep{baraffe98,chabrier00c}.
Estimated masses at a given magnitude and directly taken from the
mass-magnitude relationship extracted from the models. Samples of
masses are shown on the right-hand side of the colour-magnitude diagrams 
in Fig.\ \ref{fig_ic4665_gcs:cmd4}. We used those relations to
derive masses and construct the mass function. The highest mass cluster 
member candidate has $\sim$0.66 M$_{\odot}$ and the lowest mass object 
has 0.021 M$_{\odot}$, the latter being outside the magnitude and mass 
ranges shown in Table \ref{tab_ic4665_gcs:LF_MF_numbers}. We did not attempt 
to correct the mass function for binaries. If IC\,4665 is indeed a 
pre-main-sequence cluster, the current system mass function should 
reproduce the IMF fairly well \citep{salpeter55,miller79,scalo86}.

Figure \ref{fig_ic4665_gcs:MFplot} shows the system mass function for IC\,4665 
over the 0.66--0.02 M$_{\odot}$ mass range (filled circles and solid line). The 
mass function in logarithmic scale peaks around 0.25--0.16 M$_{\odot}$ and 
decreases down to our completeness limit at $\sim$0.035 
M$_{\odot}$ (dot-dashed line in Fig.\ \ref{fig_ic4665_gcs:MFplot}). The 
completeness in terms of mass is similar to our study in the Pleiades 
\citep[red open diamonds and solid line;][]{lodieu07c} because the difference 
in age between the Pleiades and IC\,4665 (125 Myr vs 27 Myr) is partly
compensated by the larger distance of IC\,4665 \citep{hoogerwerf01}.
The penultimate point is probably real and lies higher than the point
before at higher masses. The same trend is observed in the previous mass 
function in IC\,4665 published by \citet{deWit06} as well as in the Pleiades
mass function. \citet{moraux07a} pointed out that this rise is not caused by
an intrinsic rise of the mass function but rather reflects a drop in the
mass-luminosity relationship owing to the onset of dust in the atmospheres
of low-mass stars \citep[the M7/M8 gap as defined by][]{dobbie02b}. The 
last point of our mass function, however, suffers from 
incompleteness and represents a lower limit because the mass range 
is below our completeness limit (Table \ref{tab_ic4665_gcs:LF_MF_numbers}).
Similarly, the first point of the mass function is incomplete because of the 
saturation of the deep CFH12K images \citep{deWit06}. 

In Fig.\ \ref{fig_ic4665_gcs:MFplot} we overplotted the published mass 
function (blue open triangles and solid line) of IC\,4665 by \citet{deWit06}.
The main difference lies in the choice of the age: we have used the recent 
age estimate of $\sim$30 Myr \citep{manzi08} based on the lithium method 
\citep{rebolo92,basri96}, whereas \citet{deWit06} showed the mass function 
for 50 and 100 Myr following the earlier work by \citet{prosser93}.
We emphasise the large uncertainty on the distance of IC\,4665
\citep[385$\pm$40 pc;][]{hoogerwerf01}. This range in distance would
translate into a shift in mass with small differences in the substellar 
regime ($\sim$0.005 M$_{\odot}$) but more significant at higher masses 
(several tens of Jupiter masses for masses larger than 0.5 M$_{\odot}$).
We observe a sharper decline of the mass function derived from the
GCS data than the one derived by \citet{deWit06}, suggesting that
the level of contamination of the optical survey below 0.1 M$_{\odot}$
was higher than the one estimated by this study despite the correction
applied by \citet{deWit06}.

The mass function derived in our study is very similar to the Pleiades 
mass function drawn from the GCS dataset \citep{lodieu07c} down to 
0.05 M$_{\odot}$. 
Both mass functions peak around the same mass 
(between 0.16 M$_{\odot}$ and 0.25 M$_{\odot}$) followed by a decrease 
down to the completeness limit of the GCS survey. The difference in mass
observed at the M7/M8 gap (0.05 M$_{\odot}$ for the Pleiades vs
0.035 M$_{\odot}$ for IC\,4665) could arise from the large uncertainty on the 
distance of IC\,4665 \citep[40 pc;][]{hoogerwerf01} compared to the 5 pc 
error on the Pleiades's distance \citep{johnson57,gatewood00,southworth05}. 
Moreover, the Pleiades mass function was based on a 
photometric and proper motion sample, whereas the IC\,4665 dataset is purely 
photometric because we do not have proper motion over the full magnitude range 
probed by the GCS\@.

%
%
\section{Conclusions and future work}
\label{IC4665_GCS:conclusions}

We have presented the outcome of a wide-field near-infrared survey
of IC\,4665 conducted by the UKIDSS Galactic Clusters Survey.
The cross-matching of the GCS survey with a previous optical
survey to a similar depth leads to a revision of the membership
of low-mass stars and brown dwarfs in IC\,4665 based on their
optical-to-infrared and infrared colours. The main results of this
paper can be summarised as follows:

\begin{enumerate}
\item We confirmed a total of 493 photometric candidates identified 
in a pure optical survey as high-probability member candidates using five 
infrared ($ZYJHK$) filters in addition to the two optical ($Iz$) passbands.
\item We identified new cluster member candidates in a previously unstudied
region of the cluster.
\item We derived the luminosity and mass functions in the 
$J$ = 13.1--18.6 mag range, corresponding to masses of 0.66 and 0.02 
M$_{\odot}$ at an age of 30 Myr and a distance of 350 pc.
\item We found that the mass function is similar in shape
to the Pleiades. It is best represented
by a lognormal function peaking around 0.25--0.16 M$_{\odot}$ over
the 0.66--0.04 M$_{\odot}$ mass range.
\end{enumerate}

The full area of the cluster will be observed again in $K$ as part of the 
GCS, which will provide additional proper motion measurements over the full 
magnitude range to add proper motion information to the photometric selection. 
We are planning an optical spectroscopic follow-up with the AAOmega multi-fibre
spectrograph installed on the 3.9-m Australian Astronomical Telescope 
\citep{lewis02,sharp06} to confirm the membership of all optical and GCS
candidates. This spectroscopic follow-up will provide (i) a full spectroscopic
mass function from 0.65 to 0.04 M$_{\odot}$, (ii) a better estimate of the
level of contamination in the optical and GCS surveys, and (iii) a full
spectroscopic sequence of 30 Myr-old late-M dwarfs.
Moreover, optical and near-infrared spectroscopy of IC\,4665 members will
be extremely valuable to bridge the gap between star-forming regions like
Upper Sco \citep[5 Myr;][]{preibisch02} and older open clusters like the 
Pleiades \citep[125 Myr;][]{stauffer98} to define a temperature scale at 
30 Myr. For this reason, IC\,4665 can serve as a benchmark to study the 
evolution of the binary properties in the substellar regime, investigate the 
role of gravity as a function of age, the evolution of disks, and test
current evolutionary models. Finally, deeper surveys of IC\,4665 may reveal 
late-L and T dwarf photometric candidates of intermediate age that would be 
ideal targets for photometric and spectroscopic studies with upcoming 
facilities such as the E-ELT and James Webb Space Telescope.


%
%
\begin{acknowledgements}
NL acknowledges funding from the Spanish Ministry of Science and Innovation 
through the Ram\'on y Cajal fellowship number 08-303-01-02 and the national 
program AYA2010-19136\@. NL thanks ESO for the short but productive stay 
in May 2008 and the LAOG for his visit in December 2010\@.
We are grateful to Isabelle Baraffe for providing us with the
NextGen and DUSTY models for the CFHT/CFH12K and UKIRT/WFCAM filters.
We thank David Pinfield for his comments that improved the
content of the paper.

This research has made use of the Simbad database, operated at
the Centre de Donn\'ees Astronomiques de Strasbourg (CDS), and
of NASA's Astrophysics Data System Bibliographic Services (ADS)
and the 2MASS database.

This work is based in part on data from UKIDSS project and the CFH12K camera.
The United Kingdom Infrared Telescope is operated by the Joint Astronomy 
Centre on behalf of the U.K.\ Science Technology and Facility Council.
The Canada-France-Hawaii Telescope (CFHT) is operated by the National 
Research Council of Canada, the Institut National des Sciences de l'Univers 
of the Centre National de la Recherche Scientifique of France, and the 
University of Hawaii.
\end{acknowledgements}

%
%
  \bibliographystyle{aa}
  \bibliography{../mnemonic,../biblio_old}

%
%
\appendix

%
%
%
\section{Optical candidates confirmed by infrared photometry}
\begin{table*}
 \centering
  \caption{Coordinates, optical and near-infrared photometry, proper motions 
(in mas/yr), luminosities (in L/L${\odot}$), and mass (in M${\odot}$) for the 
493 optically selected cluster member candidates 
in IC\,4665 confirmed as such by their infrared colours extracted from the 
Eighth Data Release of the UKIDSS Galactic Clusters Survey
}
 \label{tab_ic4665_gcs:NIR_opt_cand}
 \begin{tabular}{@{\hspace{0mm}}l @{\hspace{2mm}}c @{\hspace{1mm}}c @{\hspace{2mm}}c @{\hspace{2mm}}c @{\hspace{2mm}}c @{\hspace{2mm}}c @{\hspace{2mm}}c @{\hspace{2mm}}c @{\hspace{2mm}}c c @{\hspace{3mm}}c c c@{\hspace{0mm}}}
 \hline
UGCS J\ldots{}$^{a}$ &  R.A.$^{b}$ & Dec.$^{b}$  & $I^{c}$  & $z^{c}$  &  $Z^{d}$ &  $Y^{d}$ &  $J^{d}$ &  $H^{d}$ & $K^{d}$ & $\mu_{\alpha}cos{\delta}^{e}$ & $\mu_{\delta}^{e}$ & L/L${\odot}$ & Mass \\
                     &  h m s   & $^\circ$ ' ''  & mag      &   mag    &    mag   &     mag  &    mag   &    mag   &   mag   &   mas/yr                  &    mas/yr      \\
 \hline
17:42:05.93$+$05:24:13.9 & 17:42:05.93 & +05:24:13.9 & 15.579 & 15.029 & 15.113 & 14.535 & 13.849 & 13.200 & 12.831 &    $-$1.80 &   $-$19.56 & $-$1.29 & 0.448 \\
\ldots{}             & \ldots{}    & \ldots{}    & \ldots{} & \ldots{} & \ldots{} & \ldots{} & \ldots{} & \ldots{} & \ldots{} & \ldots{} & \ldots{} & \ldots{} & \ldots{} \\
17:50:46.71$+$06:21:09.8 & 17:50:46.71 & +06:21:09.8 & 18.167 & 17.580 & 17.446 & 16.912 & 16.313 & 15.699 & 15.336 &   $-$43.42 &    25.51 & $-$2.34 & 0.084 \\
 \hline
 \end{tabular}
\begin{list}{}{}
\item[$^{a}$] Name following the IAU and UKIDSS nomenclatures
(see http://www.ukidss.org/archive/archive.html for more details).
\item[$^{b}$] Coordinates in J2000 from UKIDSS GCS DR8
\item[$^{c}$] Optical ($I$,$z$) photometry from the CFH12K survey \citep{deWit06}
\item[$^{d}$] Near-infrared ($ZYJHK$) photometry from UKIDSS GCS DR8
\item[$^{e}$] Proper motions (mas/yr) in right ascension and declination 
from the GCS vs 2MASS cross-match only for sources brighter than $J$ = 15.5 mag
\end{list}
\end{table*}
%

%
%
%
\section{Optical candidates rejected by infrared photometry}
\begin{table*}
 \centering
  \caption{Coordinates, optical and near-infrared photometry, and proper 
motions for 161 optically selected candidates in IC\,4665 classified as 
photometric non-members based on their infrared colours extracted from the 
Eighth Data Release of the UKIDSS Galactic Clusters Survey}
 \label{tab_ic4665_gcs:NIR_opt_NM}
 \begin{tabular}{@{\hspace{0mm}}l c c c c c c c c c c c@{\hspace{0mm}}}
 \hline
UGCS J\ldots{}$^{a}$ &  R.A.$^{b}$ & Dec.$^{b}$  & $I^{c}$  & $z^{c}$  &  $Z^{d}$ &  $Y^{d}$ &  $J^{d}$ &  $H^{d}$ & $K^{d}$ & $\mu_{\alpha}cos{\delta}^{e}$ & $\mu_{\delta}^{e}$  \\
                     &  h m s   & $^\circ$ ' ''  & mag      &   mag    &    mag   &     mag  &    mag   &    mag   &   mag   &   mas/yr                  &    mas/yr      \\
 \hline
174220.48$+$055253.4 & 17:42:20.48 & +05:52:53.4 & 18.925 & 18.274 & 18.389 & 17.705 & 17.043 & 16.570 & 16.172 &  $-$384.27 & 1147.56 \\
\ldots{}             & \ldots{}    & \ldots{}    & \ldots{} & \ldots{} & \ldots{} & \ldots{} & \ldots{} & \ldots{} & \ldots{} & \ldots{} & \ldots{} \\
175040.35$+$062405.9 & 17:50:40.35 & +06:24:05.9 & 18.761 & 18.066 &  ---   & 19.126 & 18.525 & 17.765 & 17.621 &   $-$54.98 & 1110.12 \\
 \hline
 \end{tabular}
\begin{list}{}{}
\item[$^{a}$] Name following the IAU and UKIDSS nomenclatures
(see http://www.ukidss.org/archive/archive.html for more details).
\item[$^{b}$] Coordinates in J2000 from UKIDSS GCS DR8
\item[$^{c}$] Optical ($I$,$z$) photometry from the CFH12K survey \citep{deWit06}
\item[$^{d}$] Near-infrared ($ZYJHK$) photometry from UKIDSS GCS DR8
\item[$^{e}$] Proper motions (mas/yr) in right ascension and declination 
from the GCS vs 2MASS cross-match only for sources brighter than $J$ = 15.5 mag
\end{list}
\end{table*}
%

%
%
%
\section{New photometric candidates identified in the GCS}
\begin{table*}
 \centering
  \caption{Coordinates and near-infrared photometry for the new cluster 
member candidates in IC\,4665 extracted from the UKIDSS GCS DR8
}
 \label{tab_ic4665_gcs:newMEMB}
 \begin{tabular}{@{\hspace{0mm}}l c c c c c c c c c c c@{\hspace{0mm}}}
 \hline
UGCS J\ldots{}$^{a}$ &  R.A.$^{b}$ & Dec.$^{b}$ & $Z^{b}$ &  $Y^{b}$ &  $J^{b}$ &  $H^{b}$ & $K^{b}$ & $\mu_{\alpha}cos{\delta}^{c}$ & $\mu_{\delta}^{c}$ \\
 \hline
UGCS J17:42:01.72$+$05:10:22.3 & 17:42:01.72 & +05:10:22.3 & 17.734 & 17.123 & 16.524 & 15.966 & 15.612 &     ---  &     ---  \\
\ldots{}                 & \ldots{}    & \ldots{}    & \ldots{} & \ldots{} & \ldots{} & \ldots{} & \ldots{} & \ldots{} & \ldots{} \\
UGCS J17:50:35.83$+$06:22:04.2 & 17:50:35.83 & +06:22:04.2 & 16.002 & 15.530 & 14.898 & 14.348 & 14.048 &   -22.14 &    17.17 \\
 \hline
 \end{tabular}
\begin{list}{}{}
\item[$^{a}$] Name following the IAU and UKIDSS nomenclatures
(see http://www.ukidss.org/archive/archive.html for more details).
\item[$^{b}$] Coordinates and photometry from UKIDSS GCS DR8
\item[$^{c}$] Proper motions (mas/yr) in right ascension and declination 
from the GCS vs 2MASS cross-match only for sources brighter than $J$ = 15.5 mag
\end{list}
\end{table*}

\end{document}